\documentstyle[]{article}
\newcommand{\bea}{\begin{eqnarray}}
\newcommand{\eea}{\end{eqnarray}}
\newcommand{\beq}{\begin{equation}}
\newcommand{\eeq}{\end{equation}}
\newcommand{\nn}{\nonumber}
\newcommand{\st}{\scriptstyle}
\newcommand{\dst}{\displaystyle}

\begin{document}
\begin{center} {\Large
On the physical meaning of Fermi coordinates}\\[1cm]
Karl-Peter Marzlin
\footnote{
Fakult\"at f\"ur Physik
der Universit\"at Konstanz,
Postfach 5560 M 674,
D-78434 Konstanz, Germany.}
\footnote{Phone: ++49-7531-883796. Electronic address:
peter@spock.physik.uni-konstanz.de}
\end{center}
\vspace{2mm}
\begin{abstract}
Fermi coordinates (FC) are supposed to be the natural extension
of Cartesian coordinates for an arbitrary moving observer in
curved space-time. Since their
construction cannot be done on the whole space and even not in
the whole past of the observer we examine which construction
principles are responsible for this effect and how they may be
modified. One proposal for a modification is made and applied
to the observer with constant acceleration in the two and four
dimensional Minkowski space. The two dimensional case has some
surprising similarities
to Kruskal space which generalize those found by Rindler for
the outer region of Kruskal space and the Rindler wedge.
In perturbational approaches the modification leads also to
different
predictions for certain physical systems. As an example we
consider atomic interferometry and derive the deviation of the
acceleration-induced phase shift from the standard result
in Fermi coordinates.
\end{abstract}
\section{Introduction}
A very convenient property of flat space-time is the existence
of Cartesian coordinates which are directly related to the
experience of an inertial observer, and which are easily
interpreted in terms of physical quantities like, e.g.,
distances. This concept should, in some sense, extend to
curved space-time since actual observers are often accelerated
and live in a (weakly) curved space. There is a general
agreement that Fermi normal coordinates are the
natural coordinate system of an observer in general space-times.
By this we mean coordinates which are most closely to
Cartesian coordinates for general relativistic situations
and which are related to observable quantities like
acceleration, rotation, or distances.
They were first introduced by M. Fermi in 1922 (Ref.
\cite{fermi22}, see also Ref. \cite{leviciv27}).
The modern construction for the freely
moving observer which is slightly different from
the original one was done by Manasse and Misner \cite{manasse63}
and was later extended to the case of an rotating and
accelerating observer by Ni and Zimmermann \cite{ni78}.
Despite of the many advantages one can take out
of FC there is unfortunately a property which inhibits that they
can have the same importance as Cartesian coordinates have:
they can't cover the whole space or even the past of the
observer. This can be seen clearly for Rindler coordinates,
the FC of an observer with constant acceleration. It is well
known that Rindler coordinates are defined only on the right
wedge of Minkowski space. The past of the origin and therefore
a part of the observers past is lost in this coordinate system.
As Mashhoon \cite{mashhoon86,mashhoon90} has discussed this
restriction means that even in classical physics
wave phenomena cannot be described correctly if the wavelength
becomes too large. It is therefore reasonable to search for
possible alternatives to FC.

Of course one is free to choose any coordinate system,
the physics does not depend on this choice. The central
supposition in this paper is that there may exist some
coordinate system which is natural in the sense described
above but which is not restricted to a neighborhood of the
worldline of the observer as it is the case for FC.
The physical interpretation of mathematical results would be
simple in this system as it is locally in FC.

In this paper we analyze the physical meaning of the
construction principles and propose a particular modification
of them. Although the modified FC do also not cover the
whole past of the observer they are less worse in this point
than Rindler coordinates. We do not claim that the modification is
better suited for a natural coordinate system. All we want to
show is that
we do not know how relatively large physical systems react on
acceleration, and that even the theoretical picture of this
case is not as clear as it might appear.
The question whether FC or (in any sense) modified FC are
natural, i.e., are directly related to measurable quantities,
is at last an experimental one.
In order to see how our modification can affect the predictions
of FC for certain experiments we have discussed the phase shift
caused by acceleration in atomic Ramsey interferometry.

Due to its simplicity the case of the Rindler observer is
suited to demonstrate the implications of our modification of
FC. We analyze it in two and four
dimensional Minkowski space. The new coordinates in two
dimensions have some surprising similarities with Kruskal space
which describes a black hole. The existence of such
similarities in the outer region (Schwarzschild space)
was shown by Rindler in 1966 \cite{rindler66}. The new feature
is that the modified FC proposed here have also resemblances
in the interior region of Kruskal space.

The paper is organized as follows. In Sec.~2 we briefly
review the construction of FC and Rindler space. The role of
the Rindler wedge is discussed. In Sec.~3 the main principle
of our proposal for the modification of FC is represented.
It is applied to the Rindler observer in two dimensions in
Sec.~4. The similarities to Kruskal space are the subject of
Sec.~5. In Sec.~6 we repeat the construction of modified FC
for the four dimensional accelerated observer. The influence
of the modification on the prediction for the acceleration
induced phase shift in atomic Ramsey interferometry is
calculated in Sec.~7.
In this paper we use the
conventions of Ref.\cite{MTW}, that is $\eta_{\mu \nu} =$
diag(-1,1,1,1), and natural units ($\hbar =c =1$).
Tetrad indices are underlined, and latin indices always run
from 1 to 3.
\section{Fermi coordinates and related problems}
In this section we briefly review the construction of FC and
discuss several properties of them. We follow the approach of
Manasse and Misner
\cite{manasse63} for a geodesically moving observer
which was later
expanded to arbitrary motion by Ni and Zimmermann
\cite{ni78}. Let $z^\mu (\tau )$ be the
worldline of the observer, where $\tau$ is his proper time which
is identical with the time coordinate of the FC (see Fig. 1).
Then $\dot{z}^\mu$
is a timelike vector normalized to -1. The dot denotes the
derivative with respect to $\tau$. At each point of the
worldline one constructs a tetrad $e_{\underline{\alpha} \mu}$
with $e_{\underline{0}}^\mu =\dot{z}^\mu$ and \cite{MTW}
\beq \frac{\dst D e_{\underline{\alpha}}}{D \tau} = -
     \hat{\Omega}\cdot e_{\underline{\alpha}}
     \label{tetdgl} \eeq
where
\beq \hat{\Omega}^{\mu \nu} = a^{\mu} u^{\nu} - a^{\nu} u^{\mu}
     + u_{\alpha} \omega_{\beta} \varepsilon^{\alpha \beta \mu
     \nu} \eeq
Here $a^\mu$ is the four-acceleration and $\omega^\mu$ the
four-rotation of the observer. In this equation of motion the
modern approach differs from the original one \cite{fermi22}
where the tetrad is simply parallel transported so that
$e_{\underline{0}}$ would in general not be tangential to the
worldline.

Any normalized tangent vector $v^\mu$ perpendicular to
$\dot{z}^\mu$ can be written as
\beq v^\mu = e^\mu_{\underline{i}} \alpha^{\underline{i}}\quad
     \mbox{ with }\sum_{i=1}^3 (\alpha^{\underline{i}})^2 =1
     \label{ortvec} \eeq

Now any point $x^\mu$ in a
neighborhood of the worldline can be connected with some point
$z^\mu (\tau_0 )$ on the worldline by a geodesic starting at
$z^\mu (\tau_0 )$ with a tangent vector perpendicular to
$\dot{z}^\mu (\tau_0 )$ (see Fig. 1).
Thus we are able to describe every point in a neighborhood
of the worldline by the proper time $\tau_0$, three variables
$\alpha^{\underline{i}}$, and the length $s$ of the geodesic
joining $x^\mu$ and $z^\mu (\tau_0)$. One can show that the
quantities $s^i := s \alpha^{\underline{i}}$ are suited for
spatial coordinates so that the FC are given by the proper time
$\tau$ and the three spatial coordinates $s^i$. Of course in
two dimensions one has only one spatial coordinate
$s^1 \equiv s$. In FC the metric of spacetime can be expanded
around the worldline ($s^i =0$) and is given by \cite{ni78}
\bea g_{00} & = & -(1 + \vec{a} \cdot \vec{s})^2 + (
     \vec{\omega} \times \vec{s})^2 - R_{0l0m}s^l s^m +
     O((s^l)^3)\nn \\ g_{0i} & = & \varepsilon_{ijk} \omega^j
     s^k - {\st \frac{2}{3}} R_{0lim} s^l s^m  \nn +
     O((s^l)^3)\\ g_{ij} & = & \delta_{ij} -{\st \frac{1}{3}}
     R_{iljm} s^l s^m + O((s^l)^3) \label{metrik} \eea
To get a feeling for the properties of Fermi coordinates it is
instructive to consider a simple example, the FC of the
constantly accelerating observer. His worldline and Zweibein
is given by \cite{MTW}
\beq z^\mu (\tau ) = \left ( \begin{array}{c} \sinh (a \tau )
     /a \\ \cosh (a\tau )/a \end{array}\right )\; ,\quad
     e^\mu_{\underline{0}} (\tau ) = \left ( \begin{array}{c}
     \cosh (a \tau )  \\ \sinh (a\tau ) \end{array}\right )\;
     ,\quad
     e^\mu_{\underline{1}} (\tau ) = \left ( \begin{array}{c}
     \sinh (a \tau )  \\ \cosh (a\tau ) \end{array}\right )
     \label{accobs} \eeq
In order to derive $e_{\underline{1}}$ in two dimensions it is
sufficient to fulfill the conditions $e_{\underline{0}} \cdot
e_{\underline{1}} =0$ and $e_{\underline{1}}\cdot e_{\underline{
1}} =1$, so that one can avoid to solve Eq. (\ref{tetdgl}).
Since the
geodesics in flat space are simply straight lines it is easy
to derive the coordinate transformation from Minkowski
coordinates to FC. This transformation is
given by $ x^\mu (\tau ,s) = z^\mu (\tau) + s
e_{\underline{1}}^\mu (\tau )$ or, written more explicitly,
\beq x^0(\tau ,s) = {(1+a s)\over a}\sinh (a\tau )\; ,\quad
     x^1(\tau ,s) = {(1+a s)\over a}\cosh (a\tau )  \eeq
The components of the metric in FC are given by
\beq g_{\tau \tau} = -(1+a s)^2\; , \quad g_{ss}=1
     \label{rindlermetrik} \eeq
which is the metric of Rindler space (see Fig. 2).

In the Rindler example one can see clearly some shortcomings of
FC when considered as the natural coordinate system of the
observer. These shortcomings are related to the fact that the
construction cannot be done in the whole space-time but only
in a neighborhood of the worldline of the observer. Although
we have not made any approximation in the construction
Rindler coordinates are
defined only in the right wedge of Minkowski space. This means
that the FC cannot even cover the whole past of the observer,
a property which would hardly be expected from his natural
coordinate system. In classical physics one may
argue that the observer becomes aware of events only if they
lie on his worldline or are carried to the worldline on light
rays. But Mashhoon \cite{mashhoon86,mashhoon90} has shown
that even classical wave phenomena are affected by the
hypothesis of locality, i.e., the local equivalence of an
accelerated observer
with a hypothetical inertial observer at the same point with
the same speed. As long as the wavelength is small compared
to $1 /a$ there are no problems. This can be related to
the Rindler coordinates by the condition $|s a| \ll 1$ which
states that the events should occur in
the immediate vicinity of the worldline.
If we leave this region we soon come into troubles: the metric
in FC becomes singular and the frequency of a wave cannot be
measured as usual \cite{mashhoon86}.
Of course those problems are even worse in quantum physics where
one usually needs a complete space-like hypersurface on which
the field is defined.

These considerations show that FC are a good tool as long as we
deal only with a small region around the worldline of the
observer. The condition $|a s| \ll 1$ is fulfilled in all
contemporary experiments. If we take an acceleration of about 10
m/s$^2$ the length $1/a$ is of the order of one light year which
is big enough for all laboratories and the solar system but too
small for stars and galaxies.

The question of interest is now whether FC are the best
coordinates which one can have to describe the experiences of
an observer. According to our supposition that there may be
one particular natural coordinate system
it may be that one can meliorate the construction
and can get rid of the unwanted restriction on a neighborhood
of the worldline while keeping most of the advantages of FC.
Despite of the fact that a coordinate system should have no
physical meaning a modification of FC, i.e., the change of the
natural coordinate system, would lead to other
predictions. The reason for this to occur is that coordinate
systems like FC
or Riemannian normal coordinates are often used
to expand physical systems in a curved space-time around a
point where the metric
becomes the Minkowski metric in these special coordinates
(see, for instance, Ref. \cite{parker79,maau93b} and references
therein).
Deviations from flat space are then growing with the distance
from this point.
A change to coordinates in which the metric reduces also
to Minkowski form at the same point but which differs from
the metric in FC in its neighborhood would
certainly change the results of the perturbational approach.
\section{Modified Fermi coordinates}
First we want to study the advantages of FC and how they are
related to the construction. One point of convenience is
the fact that the spatial coordinates $s^i$ give directly the
geodesic distance between the observers location $z^\mu (\tau )$
and some point $x^\mu (\tau , s^i)$ at the same time $\tau$.
This comes from the fact that one uses the geodesic distance to
construct the spatial coordinates.
The usage of an
orthonormal tetrad as tangents for the coordinate
lines causes the Minkowskian structure of the metric on the
worldline, a second important feature of FC.

The tetrad and its equation of motion (\ref{tetdgl}) is also
responsible for the correct description of acceleration in
the immediate vicinity of the worldline. By this we mean that
an accelerated observer sees
a test particle with mass $m$ moving in a potential
\beq V= m \vec{a} \cdot \vec{x} \label{potential} \eeq
if it is not too far away.
To show this, assume that we are close to the
worldline so that we can write
\beq g_{\mu \nu} = \eta_{\mu \nu} + h_{\mu \nu} \; , \quad
     |h_{\mu \nu}| \ll 1 \eeq
where $\eta_{\mu \nu}$ is the Minkowski metric.
It is well known that in such a situation for
non-relativistic classical and quantum test particles the main
effect of a gravitational field comes from the $h_{00}$
component and causes a term
\beq -\frac{m}{2} h_{00} \label{newton} \eeq
in the Hamiltonian
(see, for instance, Ref. \cite{maau93b} and references
therein). In principle one should also admit the case of fast
test particles but since the authors are not aware of any
experiments which measure accelerational effects
for high velocities it is omitted here.
The metric in FC is found by a coordinate transformation,
especially
\beq g_{\tau \tau}(\tau ,s^i ) = g_{\alpha \beta}(x^\mu
     (\tau ,s^i ))\; \frac{\partial x^\alpha
     (\tau ,s^i)}{\partial \tau}\; \frac{\partial x^\beta
     (\tau ,s^i)}{\partial \tau} \eeq
holds. The expansion of this equation around $s =0$
up to linear terms in $s^i$ leads to
\bea g_{\tau \tau} &\approx & g_{\alpha \beta}\frac{\partial
     x^\alpha}{\partial \tau} \frac{\partial x^\beta}{\partial
     \tau}\Big |_{s=0} + \Big \{
     \frac{\partial g_{\alpha \beta}}{\partial x^\mu} \;
     \frac{\partial x^\mu}{\partial s^i} \frac{\partial x^\alpha
     }{\partial \tau} \frac{\partial x^\beta
     }{\partial \tau} + 2 g_{\alpha \beta} \frac{\partial^2
     x^\alpha}{\partial \tau \partial s^i} \; \frac{\partial
     x^\beta}{\partial \tau}\Big \} \Big |_{s=0} s^i \nn \\
     &=& g_{\alpha \beta} e_{\underline{0}}^\alpha
     e_{\underline{0}}^\beta  + \Big \{ 2 \Gamma_{\alpha \mu
     \beta}\; e_{\underline{0}}^\alpha e_{\underline{i}}^\mu
     e_{\underline{0}}^\beta + 2 g_{\alpha \beta}\;
     e_{\underline{0}}^\beta \; \frac{d
     e_{\underline{i}}^\alpha}{d \tau}\Big \} s^i \nn \\
     &=& -1 - 2 a_{\underline{i}} s^i  \eea
where $a_{\underline{i}} = e_{\underline{i}}^\mu a_{\mu}$ are
the components of the acceleration in the tetrad frame. In the
last line use was made of Eq. (\ref{tetdgl}). Inserting this
in Eq. (\ref{newton}) leads immediately to the potential
(\ref{potential}).
It should be stressed that in the derivation we never have used
any property of FC except for the fact that the tangents of the
coordinate lines on the worldline are given by the tetrad.
Any approach which starts also with this ansatz but then uses
not geodesics but arbitrary (space-like) curves to build the
hypersurfaces of constant time has automatically the two
properties that (a) the metric on the worldline is Minkowskian,
and (b) in the limit described above an acceleration potential
of the form (\ref{potential}) arises. An observation
corresponding to (a) was already made by Levi-Civita
\cite{leviciv27} for the old formulation of Fermi coordinates.

This fact is our starting point to seek for modifications of
FC. Of course there are other trials to describe physics in
a wider range than a given neighborhood of the worldline.
Recently, Mashhoon \cite{mashhoon93} has made a phenomenological
ansatz to overcome the problem with the hypothesis of
locality. We intend to study the modifications of the metric
when some construction principles are replaced by others.
To keep the limiting potential (\ref{potential}) we also use
the tetrad as tangents of the coordinate lines.
But we replace the construction principle that the spatial
coordinates are connected with the geodesic distance.
The following observation gives the motivation for doing so.

Take one hypersurface of constant proper time $\tau$ in
FC (take, e.g., Rindler coordinates) and consider the tangent
vector $e_{\underline{i}}^{\mu}(\tau ,s^i)$
on any point of this hypersurface which is parametrized by $s^i$
(see Fig. 3). At the point $z^\mu (\tau )$ on the worldline the
vector $e_{\underline{i}}^\mu (\tau ,0)$
is perpendicular to the four velocity $e_{\underline{0}}^\mu$.
Now let us assume that the observer looks at a later time $\tau
^\prime$ at the point $x^\mu (\tau ,s^i )$. Surely his
image of the point was
transported along a light ray (null-geodesic) joining $z^\mu
(\tau^\prime )$ with the point on the hypersurface. Suppose now
he can observe the tangent vector $e_{\underline{i}}^{\mu}(\tau
,s^i)$ at this point
and that his image of this vector is given by the
parallel transportation $e_{\underline{i}}^{\prime \mu}(\tau
,s^i)$ of this vector along the light ray.
Then he will see that the spacelike direction
$e_{\underline{i}}^{\prime \mu}(\tau,s^i)$
which gives his image of the direction of the hypersurface
with constant proper time is not perpendicular to the time
direction $e_{\underline{0}}^{\mu}(\tau^\prime )$.

Of course, there is no general principle which states
that this should not be so, but neither is there a principle
which says that those hypersurfaces are build with geodesics
as for ordinary FC.
The latter construction has the appealing property that for
an inertial observer in flat space it is obvious that the FC
reduce to the usual Cartesian coordinate systems in Minkowski
space. Intuitively one would like to have the most
straightforward extension of this case when one considers
a geodesically moving observer in curved space: the straight
lines in Minkowski space should become geodesics. But this
is exactly the prescription for the construction of FC and
was shown to be valid only in a neighborhood of the worldline.

Starting from the observation made above we now propose the
following {\em construction principle} for modified FC:
The tangent vectors $e_{\underline{i}}^{\mu}(\tau
,r^i)$ at a point $x(\tau ,r^i)$ on a hypersurface of constant
proper time $\tau$ should be, after a parallel transport along
the light ray joining $x(\tau ,r^i)$ with the
worldline, perpendicular to the four-velocity
$e_{\underline{0}}^{\mu}(\tau^\prime )$ on the worldline at a
later time $\tau^\prime$.
\beq e_{\underline{i}}^{\prime \mu}(\tau,r^i)
     e_{\underline{0}\mu}(\tau^\prime ) = 0 \eeq
Here $r^i$ are variables, not connected to the geodesic
distance, which parametrize the $\tau =$ constant hypersurface.
As in the original construction we assume that they are given
by $r^i = r \alpha^{\underline{i}}$ where the $\alpha^{
\underline{i}}$
have the same meaning as in Eq. (\ref{ortvec}) and $r$
parametrizes some curve joining $z^\mu (\tau )$ and $x^\mu (
\tau ,r^i=\alpha^{\underline{i}} r)$.

Before turning to the particular case of the Rindler observer
we want to stress that this ansatz reduces to Cartesian
coordinates when we consider an observer in inertial motion in
flat space.
In flat space parallel transport is trivial so that
\beq e_{\underline{i}}^{\mu}(\tau ,r^i) =
     e_{\underline{i}}^{\prime \mu}(\tau,r^i) \label{xx} \eeq
follows. The four-velocity $e_{\underline{0}\mu}(\tau^\prime )$
of an inertial
observer is constant, i.e., it does not depend on $\tau^\prime$.
Hence $e_{\underline{i}}^{\prime \mu}(\tau,r^i)$ is constant.
Hence its parallel transportations along the light ray are
constant for all values of $\tau$ and $r^i$ on this light ray.
Since $\tau^\prime$ is completely arbitrary this holds for all
possible values of $\tau$ and $r^i$. Therefore, we can deduce
from Eq. (\ref{xx}) that the tangent vectors
$e_{\underline{i}}^{\mu}$ of the hypersurfaces of constant
$\tau$ are constant in the whole space what indicates that the
spatial coordinate lines are straight. Since the
(time-like) worldline
is also straight all coordinate lines are so. In addition,
their tangents are orthonormal per construction so that we get
a Cartesian coordinate system in Minkowski space.
\section{The Rindler observer in two dimensions}
In this section we treat the Rindler observer
in two dimensions as a clarifying example.
His worldline is given by Eq. (\ref{accobs}).
We search for a vector $e_{\underline{1}}^\prime (\tau, r)$
which is perpendicular to $e_{\underline{0}}(\tau^\prime )$.
This is easily achieved by setting $e_{\underline{1}}^{\prime}
(\tau ,r) = e_{\underline{1}}(\tau^\prime ,0)$.
To get the tangent vector $e_{\underline{1}}(\tau, r)$ on the
hypersurface with constant proper time $\tau$ we have to
parallel transport $e_{\underline{1}}^\prime (\tau, r)$
on a light ray in the past. A light ray in flat space joining
$z^\mu (\tau^\prime )$ with $x^\mu (\tau ,r)$ can be described
by
\beq x^0(\tau ,r) \pm x^1(\tau ,r)= \mbox{const.} =
     z^0(\tau^\prime
     )\pm z^1(\tau^\prime) = \pm {1\over a}e^{\pm a \tau^\prime}
     \label{lightray}\eeq
where the last equality comes from Eq. (\ref{accobs}).
The minus sign corresponds
to a light ray coming from the left side of the observers past
and going to the right side of his future.  Let us
assume the minus sign for the moment. Since the parallel
transportation in flat space is trivial we have
$e_{\underline{1}}(\tau, r)= e_{\underline{1}}^\prime (\tau,
r)$. The differential equation for the hypersurface
$\tau =$ constant is found by setting its tangent vector
$dx^\mu / dr $ equal to $e_{\underline{1}}$:
\beq \frac{d x^\mu}{dr}(\tau ,r) = e_{\underline{1}}(\tau, r)
     = e_{\underline{1}}(\tau^\prime , 0)\eeq
Using Eqs. (\ref{accobs}) and (\ref{lightray}) this becomes
\bea \frac{d x^0}{d r} &=& {1\over 2}(e^{a \bar{\tau}} -e^{-a
     \bar{\tau}}) = \frac{1}{2}\left ( a(x^0
     -x^1)-\frac{1}{a(x^0-x^1)} \right ) \nn \\
     \frac{d x^1}{dr} &=& {1\over 2}(e^{a \bar{\tau}} +e^{-a
     \bar{\tau}}) =\frac{-1}{2}\left ( a(x^0
     -x^1)+\frac{1}{a(x^0-x^1)} \right )
     \label{dgl} \eea
Since we want to
construct the modified FC with light rays in the past of the
observer we have to repeat the calculation with the plus sign
in Eq. (\ref{lightray}) in order to construct the right side
of the observers past. This leads to a similar equation like
(\ref{dgl}).
For the initial condition $x^\mu (\tau,0) = z^\mu (\tau)$
the solutions are found to be
\bea x^0 (\tau ,r) &=& \mp \frac{1}{a}e^{\mp a \tau} \pm
     \frac{1}{a}
     \cosh (a(\tau \pm r)) \nn \\
     x^1 (\tau ,r) &=& \frac{1}{a}e^{\mp a \tau}\pm \frac{1}{a}
     \sinh (a(\tau \pm r))  \label{lsg} \eea
where the sign is the same as in Eq. (\ref{lightray}).
As the plus (minus) sign refers to the right (left) side of
the past it refers to $r>0$ and $r<0$, respectively.
Figure 4 shows the lines of constant $r$
(which are always hyperbolic) and constant $\tau$
(dashed lines).
Taking Eq. (\ref{lsg}) as the coordinate transformation from
Cartesian coordinates $x^\mu$ to modified FC $(\tau ,r)$ we
find for the metric
\beq g_{\mu \nu} = \left ( \begin{array}{cc} 1-2e^{ar} &
     \mbox{sgn} (r)\, (1-e^{ar}) \\ \mbox{sgn} (r)\, (1-e^{ar})
     & 1 \end{array} \right ) \label{metrik2} \eeq
For $r=0$ the metric becomes the ordinary Minkowski metric.

Let us now see on which part of Minkowski space the new
coordinates are defined. The inverse form of Eq. (\ref{lsg}) is
\bea \tau &=& \frac{\mp 1}{a} \ln \left [ \frac{a}{2} (x^1 \mp
     x^0 ) \pm \frac{1}{2a (x^0 \pm x^1)} \right ]
     \mbox{ for } r = \pm |r| \nn \\
     r &=& \frac{1}{a} \ln \left [ \frac{1}{2}- \frac{a^2}{2}
     ((x^0)^2 -(x^1)^2) \right ] \eea
It is easy to check that, for any value of $r$, the condition
\beq \eta_{\mu \nu} x^\mu x^\nu > \pm \frac{1}{a^2}
     \mbox{ for } r = \pm |r| \eeq
holds. This means that the coordinate transformation is defined
everywhere in the right wedge and at those points in the past
and the future of the origin which lie between the two
hyperbolas described by
\beq (x^0)^2 - (x^1)^2 = \frac{1}{a^2} \eeq
Obviously our proposal for the modification of FC covers a
larger region of Minkowski space than ordinary FC. But the
main problem is not solved, i.e., the modified FC are
also not defined on the whole past of the observer.
Nevertheless, the new coordinate system uncovers new relations
between black holes and accelerated observers as will be shown
in the next section.
\section{Modified Fermi coordinates and Kruskal space}
The discovery of similarities between the accelerated observer
and black holes is not a new one. Already in 1966 Rindler
\cite{rindler66} has found that the outer or Schwarzschild
region of Kruskal space closely resembles the Rindler space.
The Kruskal space is described in most textbooks on general
relativity (see, e.g., \cite{MTW}). It was first discovered by
Kruskal \cite{kruskal60} and Szekeres \cite{szekeres60} and
describes the complete space-time of a black hole. Schwarzschild
coordinates do only cover the outer region of the hole,
and the Kruskal space is the maximal geodesically extension
of Schwarzschild space including the region inside of the
horizon. The line element is given by
\beq ds^2 = f^2 (-dv^2 + du^2) \; , \; f^2 = \frac{32m^3}{r}
     e^{-r/2m} \eeq
where we have neglected the angular part of the metric. $v$ is
the time coordinate. The radial variable $r$ of Schwarschild
space is uniquely connected to $u$ and $v$ via
\beq (\frac{r}{2m} -1)e^{r/2m} = u^2 - v^2 \label{rdef} \eeq
The coordinate time $t$ of Schwarzschild space is given by
$t = 4 m \tanh^{-1}(v/u)$ in the left and the right wedge, and
by $t = 4 m \tanh^{-1}(u/v)$ in the past and the future of the
origin.
Fig. 5 shows the lines of constant $t$ and $r$ in
Kruskal space. Note that in the inner regions the lines of
constant $r$ become space-like and those of constant $t$
time-like so that $t$ is not a proper time coordinate in the
inner region. Light rays are diagonal, the light cone of the
origin corresponds to the horizon $r=2m$ as well as to $t= \pm
\infty$. A glance at Fig. 2 shows the main results of Rindler.
In both the outer region of Kruskal space and the Rindler
wedge lines of constant time coordinate are straight and
lines of constant spatial coordinate are hyperbolic.
Qualitatively in both pictures lines of constant spatial
coordinate correspond to a uniformly accelerating rod.

A look at Fig. 4 lets suggest us that the modified FC can shed
more light on the connection between black holes and
accelerating observers. Contrariwise to the Rindler space,
lines of constant time $\tau$ are not straight but their
tangent at the worldline agrees per construction with the
tangent at the worldline in the Rindler space
and the corresponding hyperbola in Kruskal space.
The new feature is that the lines of constant $r$ are
much more closer
to the Kruskal case than those of the Rindler space.
They also are always hyperbolic, and in addition they are
defined in the part of the past of the origin which lies
above the thick dashed hyperbola. This resembles very closely
the interior region of Kruskal space-time where the hyperbola
describes the singularity. As in Kruskal space the lines of
constant spatial coordinate $r$ become space-like in the
'interior' region so that $\tau$ is here not a time coordinate.
In contrast to the Kruskal case, lines of constant $\tau$ are
still space-like (per construction) in this region so that both
$r$ and $\tau$ are not suited for the time coordinate outside
the right wedge.
\section{The four-dimensional case}
The analysis of the accelerated observer
done in two dimensional flat space can be expanded
to four dimensions. The observer is assumed to move
in the $x^1 $-direction. His worldline is given by
Eq. (\ref{accobs}) together with $z^2(\tau ) = z^3 (\tau ) =0$.
The light ray condition (\ref{lightray}) has to be replaced by
\beq (x^\mu (\tau ,r^i ) - z^\mu (\tau^\prime ))\,
     (x_\mu (\tau ,r^i ) - z_\mu (\tau^\prime )) =0
     \label{lightray2} \eeq
($i=1,2,3$). Any vector $e^{\prime}(\tau ,r^i )$
perpendicular to the observers four-velocity at the time
$\tau^\prime$ can be written as
$e^{\prime}(\tau ,r^i ) = \alpha^{\underline{i}}
e_{\underline{i}}(\tau^\prime )$ where
\beq e_{\underline{1}} = \left ( \begin{array}{c} \sinh (a
     \tau^\prime ) \\ \cosh (a \tau^\prime ) \\ 0 \\ 0
     \end{array}
     \right ) \; , \; e_{\underline{2}} = \left (
     \begin{array}{c} 0 \\ 0 \\ 1 \\ 0 \end{array} \right ) \; ,
     \; e_{\underline{3}} = \left ( \begin{array}{c}
     0 \\ 0 \\ 0 \\ 1 \end{array} \right )
     \label{tetrad} \eeq
The $\alpha^{\underline{i}}$ describe the direction of the
vector in three-space and have the same meaning as
in the ordinary construction of FC. To find an arbitrary curve
in the hypersurface of constant $\tau$ we derive the
differential equation $d x^\mu /dr = e^{\mu}(\tau ,r^i )$
with the help of Eq. (\ref{xx}) (where the index
$\underline{i}$ is dropped in the case under consideration).
$r$ is the variable
which parametrizes the curve in the hypersurface, the three
variables $r^i$ will be derived from $r$ and
$\alpha^{\underline{i}}$ as in the ordinary construction.
Inserting Eq. (\ref{tetrad}) we find
\beq \frac{d x^0}{dr} = \alpha^{\underline{1}}\sinh (a
     \tau^\prime )\; ,\; \frac{d x^1}{dr} = \alpha^{\underline{
     1}}\cosh (a \tau^\prime ) \; ,\;
     \frac{d x^2}{dr} = \alpha^{\underline{2}}\; ,\;
     \frac{d x^3}{dr} = \alpha^{\underline{3}} \eeq
where the relation between $x^\mu$ and $\tau^\prime$ is
found with the aid of Eq. (\ref{lightray2}). Again the initial
condition is given by $x^\mu (\tau ,r=0) = z^\mu (\tau )$.
The solutions $x^2$ and $x^3$ are quickly found. Introducing
\beq v^\pm := x^1 \pm x^0 \; , \quad q := 1 + a^2 (1-(\alpha
     ^{\underline{1}})^2 )r^2 + a^2 v^+ v^- \eeq
the remaining part of the equations can be written as
\beq {d v^+ \over dr} = \alpha^{\underline{1}} p \; ,\quad
     {d v^- \over dr} = {\alpha^{\underline{1}}\over p}
     \label{dgl4} \eeq
where $p$ is defined by
\beq a v^- p^2 - p q + a v^+ = 0 \eeq
This equation can be solved analytically but the solution
cannot be given in a closed form. Instead of this one can use
Eq. (\ref{dgl4}) directly to get a expansion of $x^0$ and $x^1$
in $r$. Setting
\beq r^i := \alpha^{\underline{i}} r \eeq
we find
\bea x^0 (\tau ,r^i ) &=& \frac{1}{a} \left \{ (1+a r^1) \sinh
     (a \tau ) + \frac{1}{2} a^2 r^1 r \cosh (a \tau ) +
     \frac{1}{6} a^3 r^1 (r)^2 \sinh (a \tau ) \right \} +
     O((r^i)^4) \nn \\
     x^1 (\tau ,r^i ) &=& \frac{1}{a} \left \{ (1+a r^1) \cosh
     (a \tau ) + \frac{1}{2} a^2 r^1 r \sinh (a \tau ) +
     \frac{1}{6} a^3 r^1 (r)^2 \cosh (a \tau ) \right \} +
     O((r^i)^4) \nn \\
     x^2 (\tau ,r^i ) &=& r^2 \; , \quad x^3 (\tau ,r^i ) =
     r^3 \eea
The metric in the modified FC
$(\tau ,r^i )$ is given up to order $(r^i)^2$ by
\bea g_{00} &=& -(1+a r^1)^2 \; , \quad g_{0i} = -\frac{a}{2}
     \left [ (1 + ar^1) \frac{r^1 r^i}{r} + \delta_{i 1} r
     \right ] \nn \\
     g_{ij} &=& \delta_{ij} - \frac{a^2}{4} \left ( \frac{r^1}{
     r}\right )^2 r^i r^j + \frac{a^2}{12}\left [ \delta_{i 1}
     r^1 r^j + \delta_{j 1} r^1 r^i + \delta_{i 1}\delta_{j 1}
     (r)^2 \right ] \eea
We see that $g_{00}$ has the same structure as in the
two dimensional case and that it leads to the correct potential
(\ref{potential}).
\section{Experimental significance}
Up to now we have only developed a mathematical way to
define a coordinate system which is in some way related to
the motion of the observer. The physical consequences are
less clear.
The heart of the construction scheme was the assumption of the
parallel transport of vectors along a light ray before they
can be seen by the observer. This principle is in direct
correspondence to the geodesical nature of the spatial
hypersurfaces in ordinary FC.
It may be that one of these principles is a mean to find the
natural coordinate system of the observer in which the
coordinate expressions are directly related to measurable
quantities (supposing that such a system exists). Different
candidates for the natural coordinate system can lead to
different predictions because the
relation to the measurable quantities may become erroneous with
growing distance from the worldline.

As an example we consider here the influence of the
acceleration on interferometers, especially on the atomic Ramsey
interferometer. In this device four traveling laser waves
(taken along the $x^3$ axis) serve as beam splitters
for the atomic beam which moves in the $x^1$-direction. The
details of the experiment can be
found in Ref. \cite{maau93} and the references therein
where we have calculated the first order phase shift due to a
general perturbation of the form
\beq H^{(1)} = \hat{H}(\vec{P}) \; (x^1)^{N_1}(x^2)^{N_2}
     (x^3)^{N_3} \eeq
with integer $N_i$. $\vec{P}$ is the center of mass momentum
operator of the atoms and $x^i$ the center of mass coordinates.

Consider the lowest order difference between
the $g_{\tau \tau}$ component of the Rindler metric
(\ref{rindlermetrik}) and the modified FC (\ref{metrik2}).
Expanding the latter around $r=0$ leads to
\beq g_{00} =  -(1+a r)^2 -\frac{1}{3}(ar)^3 + O(r^4 )
     \label{expansion} \eeq
The lowest order difference between the
$g_{00}$ component of Rindler coordinates and modified FC is
the third order term in Eq. (\ref{expansion}). Assuming that
the acceleration is parallel to the $x^3$-direction we
can find with Eq. (\ref{newton}) the corresponding perturbation
of the Hamiltonian:
\beq \hat{H}=  \frac{m a^3}{6} \; , \quad  N_1 =N_2 =0\; ,
\; N_3=3 \eeq
The calculation in four dimensions leads to a similar term.
Inserting this into Eq. (28) of Ref. \cite{maau93} leads to
the phase shift
\beq \Delta \varphi = -\frac{(2 \pi a T)^3 \hbar^2}{12
     \lambda^3 m^2 c^4}(2T^\prime +T) \eeq
where we have reintroduced $\hbar$ and $c$ for convenience.
$\lambda$ is the wavelength of the laser beams, $T$ and $T
^\prime$ are the times of flight of the atoms from the
first to the second laser and from the second to the third
laser, respectively. At present it should be possible to reach
times $T, T^\prime$ of about one second. As a typical value
for the wavelength we use here $\lambda$= 457 nm, the
wavelength of the intercombination transition $^3 P_1
\rightarrow \, ^1 S_0$ of $^{24}$Mg \cite{ertmer92}.
For these atoms the mass is given by $4\cdot 10^{-26}$ kg.
If $a$ is taken to be the earths acceleration $\Delta \varphi$
will be as small as $10^{-27}$, far beyond all experimental
limitations.

{}From this estimation for a typical 'local' experiment it is
clear that one has to perform experiments in accelerated
frames which include a huge
area in order to see whether modified FC or Rindler coordinates
(or anything else) describe the outcome of such an experiment
more accurately. Presently, we see no
possibility to decide this question.
\section{Conclusions}
In this paper we have addressed the question to what extent
Fermi coordinates are related to the experience of an observer
moving arbitrary in a curved space-time. The main steps in the
construction were reconsidered and applied to the case of the
observer with constant acceleration in two dimensional
Minkowski space. In this case the construction can be made
without any approximation and leads to the usual Rindler
space-time, i.e., the right wedge of Minkowski space.
The consequences of the fact that Rindler coordinates do not
cover the whole past of the observer were shortly discussed
in the context of classical physics. The physical meaning
of the different construction principles was examined.
The principle that the hypersurfaces of constant time are build
with geodesics was replaced by the supposition that those
hypersurfaces should always be perpendicular to the observers
four-velocity.

The modified construction scheme was applied
to the observer with constant acceleration in two and four
dimensional Minkowski space. The resulting modified FC covers
somewhat more of the observers past than Rindler
coordinates. In both cases there are similarities to the Kruskal
coordinates of a black hole. While the similarities in the
Rindler case are restricted to the outer region (Schwarzschild
space) the modified FC show up a new region which resembles
the interior region of Kruskal space.

The physical case of four dimensions was treated along the same
lines as in two dimensions. The resulting metric has (in an
expansion around the worldline) the same features as in two
dimensions.

Although coordinate systems do in general not have any physical
significance, the modification of FC was shown to lead to
predictions that differ from those of ordinary FC for certain
physical systems.
This is due to the fact that one can use coordinates in which
the metric becomes the Minkowski metric at certain points
to expand physics in curved space around flat space-time
physics. As a particular example we have considered atomic
Ramsey interferometry. The difference in the predictions for
the accelerational induced phase shift from FC and modified
FC is far too small to be measurable.
\\[5mm]
{\bf Acknowledgements}\\[2mm]
It is a pleasure for me to thank Rainer M\"uller for stimulating
discussions and the Studien\-stiftung des Deutschen Volkes for
financial support.

\newpage
\begin{center} \large Marzlin, On the physical meaning of
Fermi coordinates, Figure captions \end{center}
$ $\\[1cm]
Fig.1: The worldline of the observer, his tetrad and a
geodesic with tangent perpendicular to $e_{\underline{0}}$.
\\[1cm]
Fig.2: The Rindler coordinates cover only the right wedge of
Minkowski space. The thick solid line represents the worldline.
\\[1cm]
Fig.3: The vector $e_{\underline{1}}^\mu$ at apoint on the
hypersurface is, after parallel transport along a light ray,
not perpendicular to $e_{\underline{0}}^\mu (\tau^\prime )$.
\\[1cm]
Fig.4: The coordinate lines of the modified FC.
The thick solid line represents the observers worldline.
The coordinate system covers only the right wedge and the points
in the past of the origin which lie above the dashed hyperbola.
\\[1cm]
Fig.5: Kruskal space. The thick dashed hyperbolas correspond
to the singularity.

\end{document}